\documentclass[aj]{emulateapj}

\usepackage{graphicx}
\usepackage{microtype}

\usepackage{amssymb}
\usepackage{hyperref}  
\def\link_col{blue}
\hypersetup{colorlinks,citecolor=\link_col,filecolor=\link_col,linkcolor=\link_col,urlcolor=\link_col}
\hypersetup{linkcolor=red,citecolor=blue,filecolor=magenta,urlcolor=blue} 
\usepackage{xspace}
\usepackage{url}
\usepackage{wasysym}
\usepackage{journal_names}

\def \vlsr{V$_{\rm LSR}$\xspace}

\def \cmcube{\mbox{cm$^{-3}$}\xspace}
\def \cmsqr{\mbox{cm$^{-2}$}\xspace}

\def \nhthree{\mbox{NH$_{3}$}\xspace}
\def \mainCO{\mbox{$^{12}$CO}\xspace}
\def \CO{\mbox{12CO}\xspace}
\def \isoCO{\mbox{13CO}\xspace}
\def \msun{\mbox{M$_{\odot}$}\xspace}

\def \miriad{\mbox{\sc miriad}\xspace}

\def \mum{\mbox{$\mu$m}\xspace}

\def \iras{IRAS~14482--5857\xspace}
\def \pmn{PMN~1452--5910\xspace}
\def \source{PMN~1452--5910/IRAS~14482--5857\xspace}
\def\sourceIRAS{IRAS~14482--5857\xspace}
\def \water{H$_2$O\xspace}

\slugcomment{}

\shorttitle{Discovery of Luminous Star Formation in \sourceIRAS}
\shortauthors{Jones \& Braiding}

\begin{document}

\title{Discovery of Luminous Star Formation in \source: the Pterodactyl Nebula}

\author{Jones, D. I.$^1$ \& Braiding, C. R.$^2$}
\affil{$^1$ Department of Astrophysics/IMAPP, Radboud University, Heijendaalseweg 135, 6525 AJ Nijmegen, The Netherlands.}
\affil{$^2$ School of Physics, University of New South Wales, 2052, Sydney, Australia.}
\email{d.jones@astro.ru.nl}
 
\begin{abstract}
We present sensitive 1--3~GHz ATCA radio continuum observations of the hitherto unresolved star forming region known as either \iras or \pmn.
At radio continuum frequencies, this source is characterised by a ``filled-bubble'' structure reminiscent of a classical H{\sc ii} region, dominated by three point sources, and surrounded by low-surface-brightness emission out to the $\sim3'\times4'$ source extent observed at other frequencies in the literature.
The infrared emission corresponds well to the radio emission, with polycyclic aromatic hydrocarbon emission surrounding regions of hot dust towards the radio bubbles.
A bright 4.5~\mum point source is seen towards the centre of the radio source, suggesting a young stellar object. 
There is also a linear, outflow-like structure radiating brightly at 8 and 24~\mum towards the brightest peak of the radio continuum.
In order to estimate the distance to this source, we have used Mopra Southern Galactic Plane CO Survey \mainCO(1--0) and $^{13}$CO(1--0) molecular line emission data.
Integrated-intensity, velocity at peak intensity and line-fitting of the spectra all point towards the peak centred at \vlsr$=-1.1$~km/s being connected to this cloud. 
This infers a distance to this cloud of $\sim12.7$~kpc.
Assuming this distance, we estimate a column density and mass towards \sourceIRAS of $\sim1.5\times10^{21}$~\cmsqr and $2\times10^4$~\msun, implying that this source is a site of massive star formation.
Reinforcing this conclusion, our broadband spectral fitting infers dust temperatures of 19 and 110~K, emission measures for the sub-pc radio point-source of EM$\sim10^{6-7}$~pc~cm$^{-6}$, electron densities of $n_e\sim10^{3}$~\cmcube and photon ionisation rates of $N_{Ly}\sim10^{46-48}$~s$^{-1}$.
The evidence strongly suggests that \sourceIRAS is a distant -- hence intense -- site of massive star-formation.

\end{abstract}


\section{Introduction}
Massive stars are the principle source of heavy elements in the universe, and disperse material and modify their environment through winds, massive outflows, expanding H{\sc ii} regions and supernovae \citep{Zinnecker2007}.
This creates an important source of mixing and turbulence for the interstellar medium (ISM) of galaxies \citep{Zinnecker2007}, making massive star formation, and a theory for it, hugely important for modern astrophysics.
For all their effect on their host galaxies, we understand little of the conditions of massive star formation, especially in the earliest phases. 
This is because, whilst a fully developed massive star or star cluster can be observed in the optical, as well as at other wavelengths, a high amount of dust extinction during their formation makes this a demanding proposition.
Coupled with the fact that they are rare and possess short formation timescales, this makes observing the earliest phases of massive star formation in large numbers fiendishly difficult \citep{Zinnecker2007}.

Sites of massive star formation have traditionally been discovered using large surveys in the infrared, such as by using infrared colours \citep{Molinari1996,Walsh1997}, dust emission \citep{Hill2005}, or molecular line emission \citep{HOPS,Burton2013}.
However, extending the distance to which sites of massive star formation in the Galaxy can be observed is necessarily difficult, due to the large regions that must be observed to a higher sensitivity.
This is summarised by the statistic given by large surveys of mass as a function of distance that can be observed.
For instance, the \water Galactic plane survey (HOPS) can detect a 20~K, 400~\msun cloud at the $5\sigma$ level at 3.2 kpc, and a $3\times10^4$~\msun clouds anywhere in the Galaxy \citep{HOPS}.
Given this, it is becoming an imperative to use the ever-growing number of high-quality surveys across many wavebands to characterise star-forming region in a more systematic and comprehensive manner so as to increase the number of known massive star forming regions.

\begin{figure}
\centering
\includegraphics[width=0.5\textwidth]{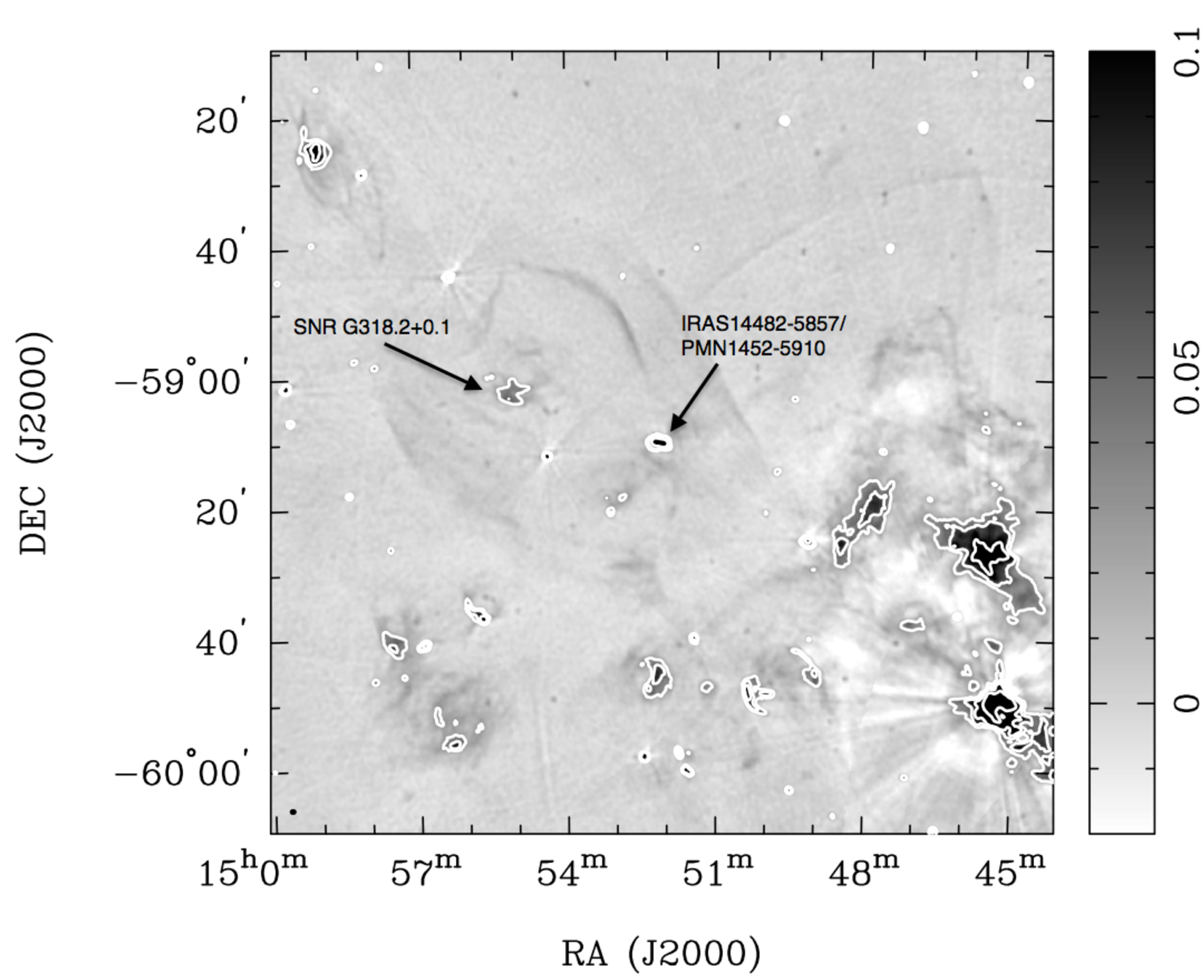}
\caption{843~MHz total intensity image of the region surrounding \sourceIRAS, overlaid with the same (white) contours at 30 (=$5\sigma$), 60, 120, 240, and 480~mJy/beam. The beam at this declination is $43''\times50''$, and is shown in the lower-lefthand corner.
}
\label{fig:SUMSS}
\end{figure}

Here we report serendipitous radio continuum observations at 1--3~GHz, as well as an examination of the archival data towards what may be a hitherto unknown site of massive star formation; \iras.
A search of SIMBAD\footnote{\url{http://simbad.u-strasbg.fr/simbad/}} within a radius of 2 arcminutes revealed only survey information for the radio (catalogued as \pmn) and infrared source (catalogued) as \iras, but nothing else\footnote{SIMBAD cites an X-ray study \citep{Xray} in reference to \pmn, but which only contains the source IRAS~14498--5856. This reference is spurious.}.
The \pmn and \iras sources are positionally coincident, and hereafter we refer to this source simply as \iras.
Figure~\ref{fig:SUMSS} shows \iras, and its surrounding environment at 843~MHz from the Molongolo Observatory Synthesis (MOST; \citealt{SUMSS_SURVEY}) Galactic Plane Survey (MGPS; \citealt{SUMSS_CATALOGUE}).
\iras lies near the large ($40'\times35'$) supernova remnant, G318.2+0.1, along the Galactic plane at a position of $\alpha=14^{\textrm{h}}52^{\textrm{m}}05\fs3$, $\delta=-59^\circ10'08\farcs58$~(J2000.0).

\section{New Observations and Archival Data}
In this section we describe the new data on \iras that was obtained using the ATCA, as well as the archival data obtained during a literature search, as well as survey data that we possessed.

\subsection{New Australia Telescope Compact Array Data}\label{sec:data}
The ATCA data presented here are a part of an observational campaign towards a nearby object, but \sourceIRAS is contained within the field of view at 1--3~GHz.
The observations and data reduction technique will be detailed elsewhere, but we give a brief overview here.

A (hexagonal) mosaic observation of the region were performed with the ATCA on the 5$^{\textrm{th}}$ May and 6$^{\textrm{th}}$ September 2013 in the 6.0C and 1.5A array configurations, respectively.
Both array configurations have a minimum baseline of 153 meters, up to a maximum of 6000 meters (6.0C configuration only -- the maximum 1.5A baseline is 4469 meters).
The observations used the 16cm receiver with the CFB 1M-0.5k correlator, which gives a 2~GHz bandwidth with 2048 channels of 1~MHz width over the 1--3~GHz band.
We split up the band into four 512~MHz-wide sub-bands, and use only the 1332, 1844 and 2868~MHz bands here.
Standard calibration and imaging methods were used within the \miriad software reduction package.
The flux density calibrator used was PKS~B1934--638, bootstrapped to a flux density of $11.986\pm0.005$~Jy/beam at 2100~MHz.
These bands have a resolution of $10.6''\times6.8''$, $\sim8.6''\times5.4''$ and $5.1''\times3.4''$ and $1\sigma$~r.m.s. sensitivities of $\sim400$, $\sim200$ and $\sim100$~$\mu$Jy/beam at 1332, 1844 and 2868~MHz, respectively.
We found a systematic uncertainty of $\sim5$\% for the observations and all errors quoted in this paper use this added in quadrature to the local r.m.s. sensitivity.
It is important to note that these observations will recover spatial information only up to $\sim5.5'$, $\sim4'$ and $\sim2.7'$ at 1332, 1844 and 2868~MHz.
Given that the source size is $\sim3'\times4'$, the upper band may not fully recover the entire flux density.

\begin{figure*}
\centering
\includegraphics[width=\textwidth]{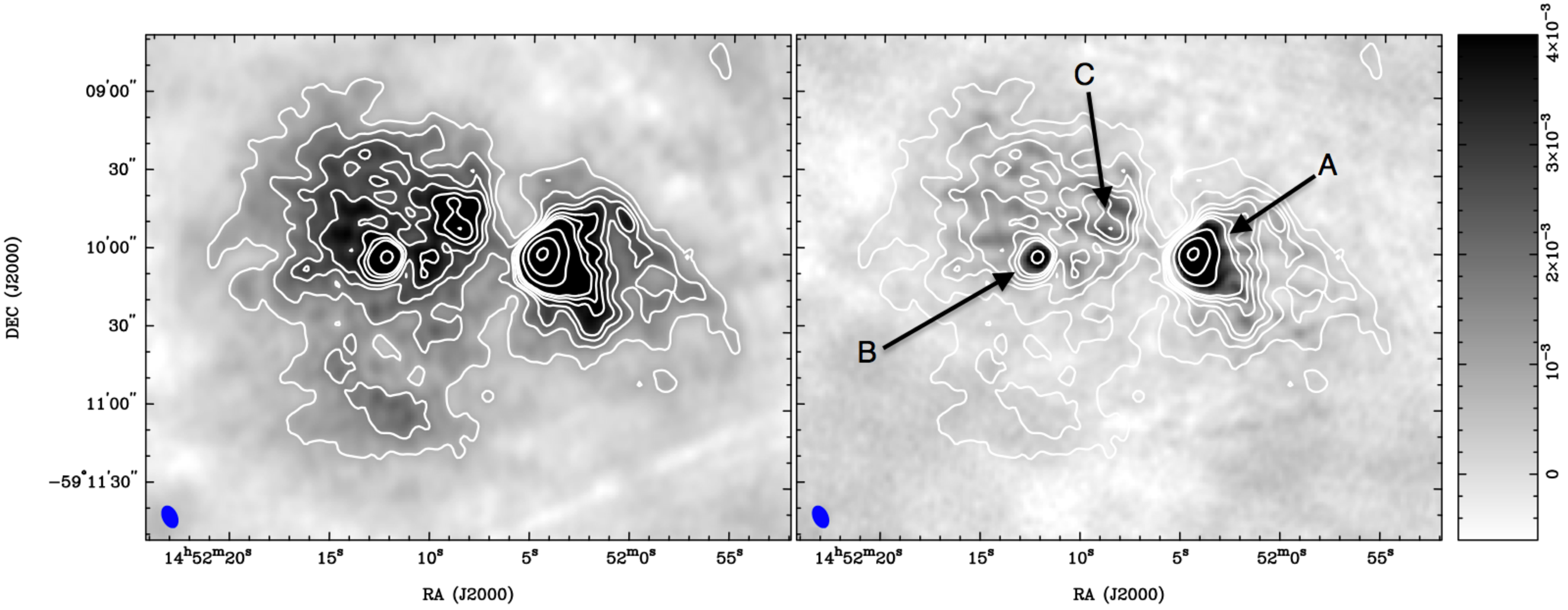}
\caption{Total intensity images of our ATCA data at 1844 (left) and 2868 (right) of \sourceIRAS. 
The contours in both frames are 1844~MHz contours at 0.8, 1, 2, 4, 6, 7, 8, 10, 20, 40, and 80~mJy/beam. 
The intensity scale is linear and runs from -0.6 to 4~mJy/beam. At 1844~MHz, the r.m.s. noise is $\sim200$~$\mu$Jy/beam (so that 0.8~mJy/beam $=4\sigma$), whereas at 2868~MHz, the r.m.s. noise is $\sim100$~$\mu$Jy/beam.
The beams are $8.6''\times5.4''$ and a position angle of $25.1^\circ$ and $5.1''\times3.4''$ at a position angle of $21.6^\circ$ at 1844 and 2868~MHz respectively. 
The beam are shown in the lower-lefthand corner of both images.
The regions that are peaks in intensity are labeled `A', `B', and `C' in the righthand panel.
}
\label{fig:ATCA}
\end{figure*}

\subsection{Archival Data}
In order to better characterise the \iras source, we have performed a literature search, as well as obtained data from several surveys.
Here we describe the data products we have obtained and used.

\subsubsection{Millimeter \& Infrared Data}
Since this source is identified in SIMBAD as \iras, we know that it is detected at infrared frequencies.
Hence we have obtained mid-infrared data from Infrared Astronomical Satellite (IRAS) at 12, 25, 60 and 100~\mum.
We have also obtained IRAC Mid-course Space eXperiment (MSX) and {\it Spitzer} Galactic Legacy Infrared Mid-Plane Survey Extraordinaire (GLIMPSE) data at the shorter wavelengths of 8.28, 12.13, 14.65, and 21.34~\mum for the A, C, D and E bands of MSX and at 3.6, 4.5, 5.8 and 8~\mum for the GLIMPSE bands. 
Emission at 3.6~\mum is dominated by stellar emission, and is usually represented in 3-colour (i.e., false-colour) images as blue.
Bright emission at 4.5~\mum is thought to be due to shocks arising from outflows from protostellar objects interacting with the ambient ISM \citep{Cyganowski2008b}.
5.8 and 8.0~\mum emission is thought to be due to polycyclic aromatic hydrocarbons (PAHs), which are particularly useful for highlighting dust emission, since they are very sensitive to infrared emission \citep{Churchwell2009}.
We have also made use of the longer wavelength MIPSGAL 24~\mum emission, which also traces dust emission.
Finally, we have searched archival papers for information on dust clumps observed at 1.1\,mm (150~GHz), which is also thought to be a powerful discriminator of the evolutionary stage of massive star formation (see, for instance, \citealt{Breen2010}).

\subsubsection{Mopra $^{12}$CO(1--0) and $^{13}$CO(1--0) data}
In order to establish the presence (or absence) of molecular gas towards \iras, we have obtained molecular line emission data from the Mopra telescope from the Mopra Southern Galactic Plane CO Survey \citep{Burton2013}.
This survey used the Mopra telescope, which is a single 22\,m dish radio telescope located $\sim450$~km north-west of Sydney, Australia ($31^\circ16'04''$S, $149^\circ05'59''$E, 866-m a.s.l.) to survey a region of the southern Galactic plane from $l=305^\circ - 345^\circ$ and $b=\pm0.5^\circ$.
The survey observations utilised the UNSW Mopra wide-band spectrometer (MOPS) in zoom mode with the 3\,mm receiver, which operates in a frequency range of 77--117~GHz, and for the relevant observations here, the 8~GHz-wide bandwidth was centred on 112.5~GHz.
The fast-mapping zoom mode of MOPS allows observations in up to 8 windows simultaneously, where each window is 137.5 MHz wide and contains 4096 channels in each of two polarisations.
This observational set-up yields a spectral resolution of $\sim0.09$~km/s over a velocity range of $\sim-500$ to 500~km/s depending on the line sampled, and a spatial resolution of $\sim35''$, with the main beam efficiency of 0.55 at 115~GHz. 
The observations were conducted using the `fast on-the-fly' (FOTF) method, and reduced as per the methods described in \cite{Burton2013}.
This resulted in a $1\sigma$ sensitivity for the $^{12}$CO and $^{13}$CO (hereafter simply \CO and \isoCO) lines of $\sim1.5$~K and $\sim0.7$~K per 0.1~km/s velocity channel, respectively.

\subsubsection{\nhthree(1,1) and (2,2) inversion transition and H$_2$O maser emission}\label{sec:HOPSdata}
As presaged above, it is thought that massive stars are formed in dense clumps of cold gas.
Since CO is known to not be a particularly good tracer of cold, dense molecular material because it ``freezes out'' in temperatures of 10s of K \citep{Nicholas2011}, as well as suffering optical depth problems at moderate densities, we have obtained data from the H$_2$O Galactic Plane Survey (HOPS; \citealt{HOPS}).

HOPS is a survey of $\sim100^\circ$ region of the Galactic plane (from $-70^\circ>l>30^\circ$, $|b|<0.5^\circ$) using the Mopra telescope in the 12-mm waveband in on-the-fly mode also using MOPS.
HOPS is particularly useful for tracing massive star formation because in a single survey, it traces the 23~GHz water maser transition that is known to trace star formation, as well as the versatile inversion transitions of the ammonia molecule (which do not suffer from freeze-out, and trace cold, dense clumps).
The \nhthree(1,1) and (2,2) data have a resolution of $2'$ over velocities ranging from $\pm200$~km/s.
The median sensitivity of the \nhthree data cubes is $\sigma_{T_{mb}}=0.20\pm0.06$~K. 

\section{Results}\label{sec:Results}
\subsection{Radio Continuum Emission}
\begin{figure}
\centering
\includegraphics[width=0.5\textwidth]{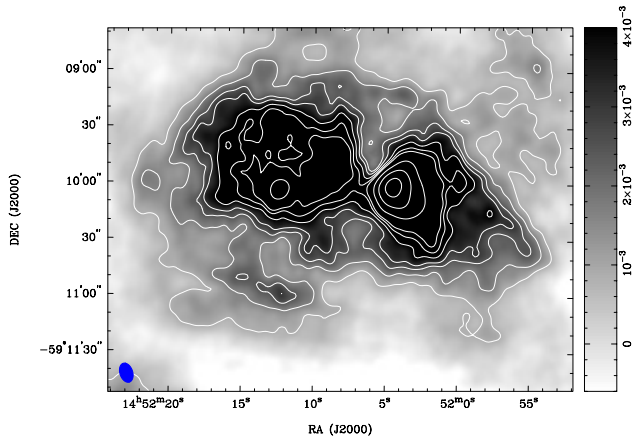}
\caption{Total intensity image of \iras at 1332~MHz. 
The (white) contours are at levels of 0.2, 0.4, 0.8, 1, 2, 4, 8, 16, 32, and 64~mJy/beam. 
The intensity scale is linear and runs from -2 to 6~mJy/beam. 
The r.m.s. noise is $\sim400$~$\mu$Jy/beam (first contour is 0.8~mJy/beam $\sim2\sigma$). 
The beam is $10.6''\times6.8''$ and a position angle of $18.8^\circ$ and is shown in the lower-lefthand corner.
}
\label{fig:ATCA_1332}
\end{figure}

Figure \ref{fig:ATCA} shows total intensity radio continuum images of \iras at 1844 and 2868~MHz.
The source exhibits a bi-lobal structure, consisting of diffuse, low-surface-brightness emission, as well as at least three strong peaks, with a single peak seen in the western lobe of the bubbles, and two in the eastern one.
The lack of diffuse emission at 2868~MHz in Figure~\ref{fig:ATCA} (right), shows the results of the spatial filtering of interferometers.
However, we suggest that the difference between the true flux density distribution of the source and that shown here is not significant.
This is because, as is shown in Table~\ref{table:PSs}, the integrated flux densities of the strong point sources are consistent with optically-thin thermal emission, and match those seen at lower (843~MHz) and higher (4500~MHz) frequencies.
Spectral index maps (see below) also show optically-thin thermal emission over most of the bubbles, suggesting that significant amounts of flux at 2868~MHz have not been filtered.
At 1332~MHz, as Figure~\ref{fig:ATCA_1332} shows, the opposite is true.
This figure shows \iras at 1332~MHz, and illustrates that there is ample diffuse emission at this frequency, since it samples structures larger than the source itself.
Indeed, the larger source extent, along with an integrated flux density for the entire source of $\sim2$~Jy suggests that at this frequency, a significant amount of background emission from the Galactic plane is being sampled.
Hence we do not use this datum in fitting the centimetre radio continuum emission.
Again -- although in the opposite sense -- the flux density reported from the PMN survey at 4500~MHz (using single dishes) is commensurate with the 1844 and 2868~MHz flux densities shown in Table~\ref{table:PSs}.
Hence we conclude that the additional flux density in the 1332~MHz is background emission, rather than reflecting additional flux density from the source itself.

\begin{figure}
\centering
\includegraphics[width=0.52\textwidth]{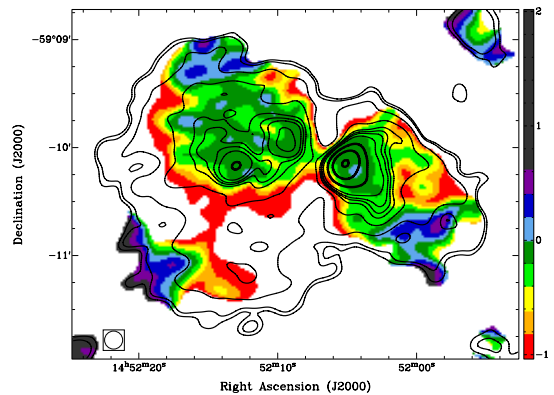}
\caption{
Spectral index map of \iras between 1844 and 2868~MHz at a common resolution of $10''\times10''$.
The (black) contours are 1844~MHz emission at 0.8, 1, 2, 3, 4, 5, 6, 7, 8, 12 and 16~mJy/beam.
This illustrates the non-thermal nature of parts of the radio bubbles.
The peaks of intensity within the radio bubbles show a thermal structure, commensurate with the total integrated flux.
}
\label{fig:SPIN}
\end{figure}

Figure~\ref{fig:SPIN} shows a spectral index map of \iras between 1844 and 2868~MHz.
We define the spectral index $\alpha$ between two frequencies $\nu$ as $\alpha\equiv d\log S_\nu/d\log\nu$.
Figure~\ref{fig:SPIN} shows the spectral index variations between $-1$ (indicative of non-thermal emission, e.g., synchrotron emission) and $+2$ (indicative of thermal emission, e.g. optically thick thermal emission).
This indeed illustrates the point made above, that \iras consists of optically thin thermal emission, interspersed with some non-thermal emission towards the edge of the radio bubbles, which is most likely due to low signal-to-noise in the 2868~MHz data.
The same map between 1332 and 1844~MHz shows a similar spectro-morphological structure to that shown in Figure~\ref{fig:SPIN}, but with steeper spectral indices, suggestive of contaminating flux density.

We have characterised the strong peaks observed at 1332, 1844 and 2868~MHz, and labelled as sources A, B and C in the right-hand panel of Figure~\ref{fig:ATCA}.
This was done by fitting Gaussian sources to the data using the \miriad task {\it imsad}, without fitting a background to subtract.
We do not subtract any background emission because, as shown above, the size scales that these observations trace are not larger than the source, and hence we would only be subtracting source flux density.
A more careful analysis would take this into account at 1332~MHz, however.
Table~\ref{table:PSs} shows the results of this fitting, and shows the source name (column 1), frequency fitted (column 2), best-fit positions (columns 3 and 4), angular sizes and position angles (column 5) of the best-fit Gaussians, as well as the peak and integrated flux density (columns 6 and 7).
As can be seen in Figure~\ref{fig:ATCA} (right), the second peak in the eastern bubble is not detected at 2868~MHz.
The diffuse flux density has average values of 6, 3 and 3~mJy/beam at 1332, 1844 and 2868~MHz respectively, with the stronger average flux density at 1332~MHz probably due to the background flux density at scales larger than the source.
These levels were used as a cut-off level for the modelling of the bright radio peaks.

\begin{deluxetable*}{cccclrr}
\tabletypesize{\scriptsize}
\tablecaption{Observed parameters of the sources within \iras at 1332, 1844 and 2868~MHz as marked in Figure~\ref{fig:ATCA}.\label{table:PSs}}
\tablewidth{0pt}
\tablehead{
\colhead{Source} & \colhead{Frequency} & \colhead{$\alpha$} & \colhead{$\delta$}  & 
\colhead{Angular Size/P.A.} & \colhead{$I_\nu$} & \colhead{$S_\nu$} \\
\colhead{} & \colhead{(MHz)} & \colhead{(J2000.0)} & \colhead{(J2000.0)} & \colhead{($'',^\circ$)} &
\colhead{(mJy/beam)} & \colhead{(mJy)} \\
\colhead{(1)} & \colhead{(2)} & \colhead{(3)} & \colhead{(4)} & \colhead{(5)} & \colhead{(6)} & \colhead{(7)} 
} 
\startdata
A & 1332 & 14:52:04.26 & -59:10:03.0 & $20.4\times18.5$, -5.1 & $51.9\pm0.3$ & $272\pm13$ \\ 
   & 1844 & 14:52:04.34 & -59:10:01.9 & $15.6\times13.1$, -15.8 & $45.2\pm0.2$ & $200\pm10$ \\
   & 2868 & 14:52:04.35 & -59:10:01.5 & $11.9\times10.0$, -16.0 & $23.0\pm1.0$ & $155\pm8$ \\
B & 1332 & 14:52:11.06 & -59:09:56.8 & $106.1\times48.9$, -71.1 & $8.4\pm0.4$ & $606\pm30$ \\
   & 1844 & 14:52:11.51 & -59:09:56.8 & $90.6\times44.1$, -69.8 & $4.8\pm0.2$ & $418\pm21$ \\
   & 2868 & 14:52:12.25 & -59:10:02.8 & $07.9\times07.3$, -46.7 & $7.7\pm0.4$ & $25\pm1$ \\	
C & 1332 & 14:52:13.20 & -59:09:46.4 & $43.1\times22.6$, 1.9 & $6.1\pm0.3$ & $82\pm4$ \\
    & 1844 & 14:52:00.37 & -59:09:47.7 & $20.1\times10.0$, +29.0 & $3.6\pm0.2$ & $16\pm1$ \\
\enddata
\tablecomments{Units of right ascension are hours, minutes and seconds, and units of declination are degrees, minutes and seconds.}
\end{deluxetable*}

\subsection{Archival Data}
\subsubsection{Radio Continuum and Millimetre Data}\label{sec:archivalRadio}
Data associated with \pmn comes from the Parkes-MIT-NRAO survey (PMN; \citealt{PMN}) and comprises a flux density and angular extent of $1.14\pm0.06$~Jy and $3'\times4'$ at a position angle of $128.5^\circ$, respectively.

A SIMBAD search of \iras contains flux densities from a coincident source found using the QUaD telescope from their Galactic plane survey \citep{QUADsurveyA}, which lists a flux density of $2.35\pm0.44$ and $1.90\pm0.3$~Jy at 100 and 150~GHz, respectively \citep{QUaDsurveyB}.
This flux density comes from a source extent of $3.9'\times3.1'$ and $2.8'\times1.7'$ at 100 and 150~GHz respectively, roughly matching that found at centimetre wavelengths (i.e., the PMN survey and the extent of the source in Figure~\ref{fig:SUMSS}).
\citet{QUaDsurveyB} also characterise this source as an UCH{\sc ii} region based on the Wood-Churchwell criteria which is based on the IR colours \citep{Wood1989}.

\subsubsection{Mopra CO(1--0) Molecular Line Emission}\label{sec:CO}
\begin{figure}
\centering
\includegraphics[width=0.5\textwidth]{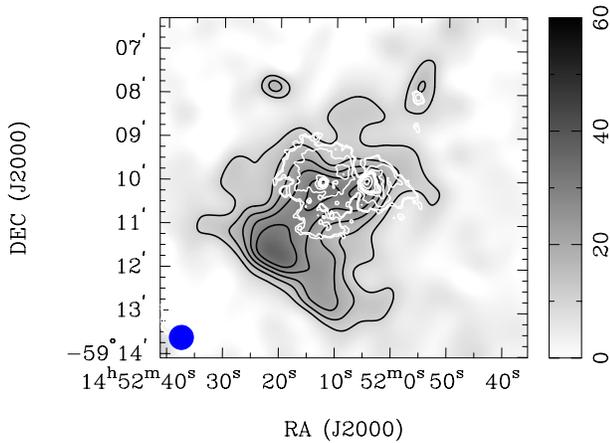} 
\caption{Integrated-intensity image of the region surrounding \sourceIRAS obtained using the velocity-cube for the \CO-line and integrated over the velocity range of -10 to +10~km/s, overlaid with the same contours at 10, 15, 20, 25, 30 and 35~K~km/s (the peak emission is at 35~K~km/s).
Overlaid on the integrated-intensity image are (white) contours of our 1844~MHz ATCA data, using the same contours as in Figure~\ref{fig:ATCA}.
The intensity range (for a linear scaling) runs from 0 to 60~K~km/s, so as to illustrate the peaks of emission coinciding with the radio continuum peaks.
The $35''$ beam of the \CO emission is shown in lower left-hand corner.
}
\label{fig:CO_mom0}
\end{figure}

Figure~\ref{fig:CO_mom0} shows an integrated-intensity (i.e., $mom0$) image for the \CO data integrated over the velocity range of -10 to +10~km/s.
This velocity-range is motivated by inspection of velocity-at-peak-intensity images (not shown here) derived for this region that show that the peak of emission within the (white) contours of Figure~\ref{fig:CO_mom0}, as well as the more extended emission, lie between this velocity range.
Integrating the \CO emission between -55 and -35~km/s (images also not shown here) does not reveal emission that follows the same morphology as Figure~\ref{fig:CO_mom0}.
This suggests that the material located near 0~km/s is associated with \sourceIRAS.
On the other hand, Figure~\ref{fig:CO_mom0} shows the positional coincidence of the peak of emission of both the \CO and radio continuum (peaks A and B of Figure~\ref{fig:ATCA}).
This Figure also illustrates the extension of the molecular emission to the southeast of the radio source, and its termination in a linear-like feature running roughly northeast to southwest and represented by the tight grouping of the contours found there.

\begin{figure}
\centering
\includegraphics[width=0.5\textwidth]{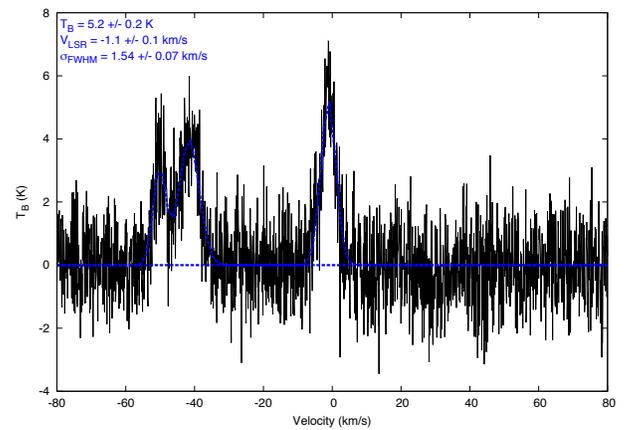}
\caption{Brightness temperature versus velocity plot for the \CO emission towards the peak of the emission found towards \sourceIRAS, as shown in Figure~\ref{fig:CO_mom0}.
The derived parameters for the fit to the data (as shown by the -- blue -- dashed line) for peaked emission at $-1.12 \pm0.1$~km/s are shown in the top left-hand corner of the plot.
}
\label{fig:CO_line}
\end{figure}

Figure~\ref{fig:CO_line} shows the line profile for the \CO emission towards the peak of the radio emission at 1844~MHz, given as source~A in Figure~\ref{fig:ATCA} ({\it right}).
We have fit (using an iterative $\chi^2$ fitting procedure) the resulting spectrum for this position and find that a two-plus-one component fit represents the data best.
We fit a two-component Gaussian to the $\sim-42$~km/s line emission, one at $V_{LSR}=-41.6\pm0.2$~km/s with a brightness temperature of $T_{mb}=4.0\pm0.1$~K and a full-width at half-maximum (FWHM) of $\sigma_{FWHM}=2.1\pm0.1$~km/s.
The second peak is fit using a brightness temperature of $T_{mb}\sim3.1\pm0.2$~K at a velocity of $V_{LSR}=-50.2\pm0.2$~km/s at a FWHM of $\sigma_{FWHM}=1.3\pm0.1$~km/s.
The peak of emission at $\sim0$~km/s is fit using a brightness temperature of $T_{mb}=5.2\pm0.2$~K, at a velocity of $V_{LSR}=-1.1\pm0.1$~km/s and a FWHM width of $\sigma_{FWHM}=1.5\pm0.1$~km/s.
This fitting shows that, in addition to the peak-velocity analysis described above, the emission component at $V_{LSR}=-1.1\pm0.1$~km/s is stronger than the $V_{LSR}=-41.6,-50.2$~km/s components.
Using the \isoCO data cube, we have similarly fit the emission from the peak of the radio continuum.
Although the weaker nature of the lines here means that we could not reliably fit any emission around the $-42$~km/s line seen in the \CO cube, we could fit a component to the emission around $-1.1$~km/s.
Doing this results in parameters of $T_{mb}\sim1.4\pm0.1$~K at a velocity centroid of $V_{LSR}=-0.9\pm0.1$~km/s and a FWHM width of $\sigma_{FWHM}=1.0\pm0.1$~km/s.
Using the velocity at peak intensity for each of the components in the \CO data, and assuming that \iras is indeed associated with these clouds, we derive near (far) distance estimates of $\sim3.5$(9.0)~kpc and 2.85(9.75)~kpc for the $-50.2$ and $-41.6$~km/s peaks respectively and $\sim12.7$~kpc for the $-1.1$~km/s peak.

\subsubsection{\water Maser and Ammonia Line Emission}\label{sec:water}
From the HOPS data and catalogue \citep{HOPS}, we find that a water maser source with a peak flux density of 5.1~Jy at -3.3~km/s with a line-width of 1.0~km/s is catalogued at a position of (R.A., Dec.; J2000.0)=$14^{h}52^m14.5^s,-59^\circ10^m49.0^s$.
From the given conversion factor of 12.5 at 22.2~GHz, this flux corresponds to a main beam brightness temperature of 0.4~K.

\begin{figure}
\centering
\includegraphics[width=0.5\textwidth]{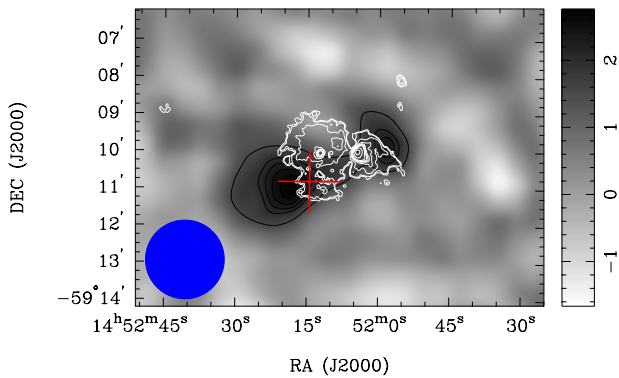}
\caption{Velocity-integrated image of 22.2~GHz \water maser emission surrounding \iras.
The emission seen here has been integrated between -10 and +10~km/s, and is overlaid with the same (black) contours at 1.5, 2, 2.2, 2.4, and 2.6~K.
The (red) cross represents the maser emission reported by the HOPS collaboration \citep{HOPS}.
The (white) contours are 1844~MHz radio continuum emission at 0.8, 1, 2, 4, 8 and 16~mJy/beam.
The $2'$ Mopra beam for the HOPS data is shown in the lower left-hand corner.
}
\label{fig:H20maser}
\end{figure}

Figure~\ref{fig:H20maser} shows the velocity-integrated emission (integrated between -10 and +10~km/s) towards \iras.
In addition to the position of the maser emission reported in \citet{HOPS}, which is described by the (red) cross, it shows two regions of possible maser emission.
The first peak (at $\sim2.7$~K\,km/s) is located at $\alpha=14^{\textrm{h}}52^{\textrm{m}}17\fs8$, $\delta=-59^\circ10'50\farcs2$~(J2000.0), and is the maser emission reported above.
The second peak (at $\sim2.3$~K\,km/s) is located at $\alpha=14^{\textrm{h}}51^{\textrm{m}}58\fs3$, $\delta=-59^\circ09'50\farcs2$~(J2000.0), which is located $\sim2.7'$ angular distance from the reported maser spot, which is a separation of about 1.2 beams.
Thus, we take only the listed brightness temperature, velocity and velocity width listed in \citet{HOPS}, suggesting that observations with a higher angular resolution instrument such as the ATCA is required to reveal the nature of the \water maser emission throughout \iras.

Using a velocity centroid of $v=-3.3$~km/s, and assuming that this source is associated with the radio emission, we arrive at a distance to \iras of 12.5~kpc.
This is similar to the distance derived above from the \CO emission, and hereafter, we adopt a distance to \iras of 12.5~kpc. 
This implies that the source extent results in an intrinsic source size of $\sim11\times14.5$~pc.

Utilising the publicly available data cubes from the HOPS website\footnote{This can be found at: \url{http://awalsh.ivec.org/hops/public/index.php}}, we have searched for emission from the inversion transitions of the \nhthree(1,1), (2,2) and (3,3) lines.
We found no emission coincident with the \iras source at either velocity down to the limiting $\sigma_{T_{mb}}=0.20\pm0.06$~K.
For the near distance, HOPS would observe a 400~M$_\odot$ cloud here with a $5\sigma$ level (whereas this increases to $\sim10^4$~M$_\odot$ at the far distance).
Interestingly, for the (1,1) transition, this equates to a 3.2 kpc distance limit for detecting a 20~K, 400~M$_\odot$ cloud at the $5\sigma$ level, with more massive clouds being observable further away with a $3\times10^4$~M$_\odot$ cloud being observed anywhere in the Galaxy.
We note that even though \water maser emission has been known to appear at velocities far from the systemic velocity of the source that created it \citep{Breen2010b}, the strong \CO lines fitted in the previous section suggest that this is not the case here.

\subsubsection{Other Masing Emission Surveys}\label{sec:otherMasers}
Maser searches, particularly the 6.7~GHz methanol maser line first detected by \cite{Menten1991} over 20 years ago, are excellent discriminators of massive star forming regions.
Indeed, it has recently been shown that this masing transition is exclusively associated with massive star formation \citep{Breen2013}.
However, surveys searching for 6.7~GHz methanol masers have not detected any significant maser emission from the region down to a limiting flux density of 0.3~Jy, for a beam size of $3.3'$ \citep{Walsh1997}.
We have searched the literature for other observations of this region using other maser tracers, class~{\sc i} methanol masers at 33 and 44~GHz \citep{Voronkov2014}.
No survey, other than HOPS, has reported maser emission from \iras (i.e., the methanol maser surveys -- e.g. \citealt{Caswell1998,Green2012}).

\subsection{Infrared Emission}\label{sec:FIR}
\begin{figure*}
\centering
\includegraphics[width=\textwidth]{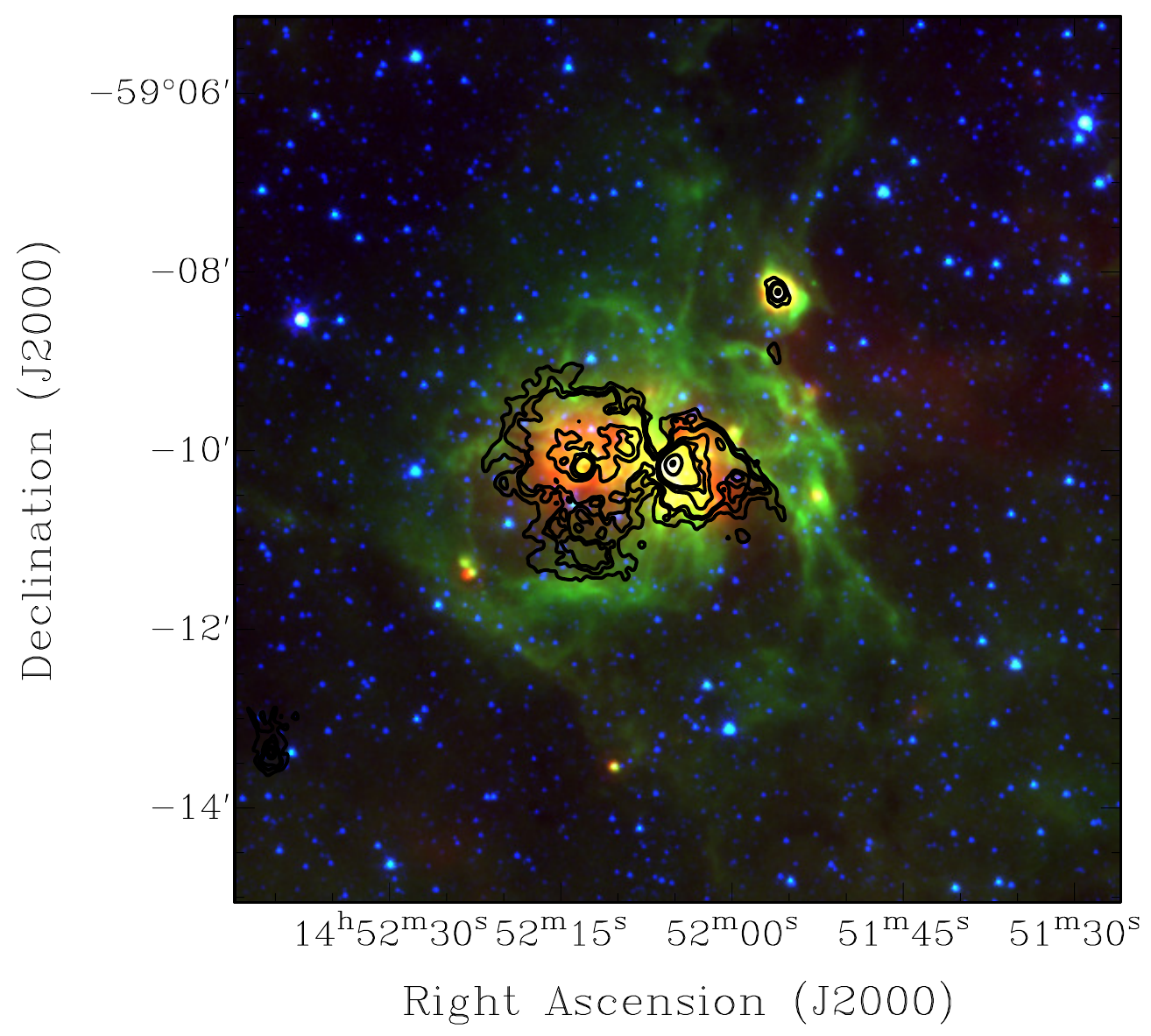}
\caption{A 3-colour image of the region surrounding \iras obtained using 3.6~\mum (blue), 8~\mum (green) data from the GLIMPSE survey, as well as 24~\mum (red) data from MIPS.
The 24~\mum emission runs from 6 to 162~MJy/sr, whilst the 3.6 and 8~\mum emission runs from 0.3 to 24.5 and 27 to 127~MJy/sr, respectively under a linear transfer function.
The contours, now black and at levels starting at 0.8~mJy/beam increase as $0.8*2^n$ for $n\in[1,8]$, in order to maximise clarity in the 3-colour image.
}
\label{fig:MSX}
\end{figure*}

Figure~\ref{fig:MSX} presents a 3-colour image of \iras at 3.6 (blue; GLIMPSE/IRAC), 8 (green; GLIMPSE/IRAC) and 24~\mum (red; MSX/MIPS).

This image shows a complex source morphology.
The medium surrounding \iras is dominated by 8~\mum emission, which is sensitive to polycyclic aromatic hydrocarbon (PAH) emission, and a strong tracer of massive star formation \citep{PAHs2008}.
This PAH emission is wispy and filamentary in nature, extending some $\sim1-2'$ further than, and appears to confine, the radio emission.
The extension of the diffuse radio continuum emission towards the southwest and southeast of the source follows minima of the PAH emission.
Hot dust emission, traced by the 24~\mum emission, is most obviously seen within the diffuse radio filled bubble morphology of a classical H{\sc ii} region of this source, although this is mixed with significant amounts of PAH emission, giving large parts of the bubbles the yellow colour seen in the image.
The southwest bubble extends further out, and matches low-surface brightness emission traced by hot dust emission.

\begin{figure}
\centering
\includegraphics[width=0.5\textwidth]{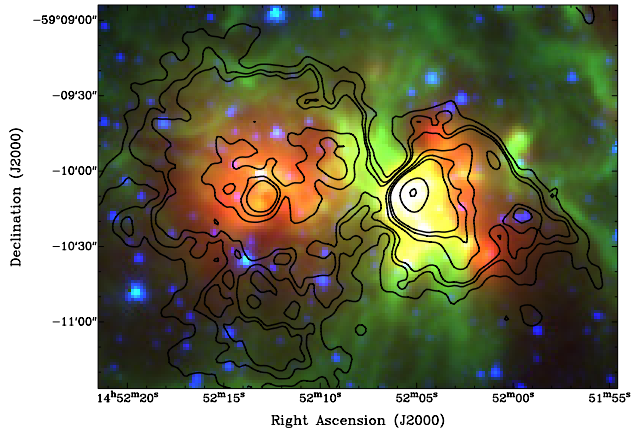}
\caption{Closeup of the diffuse radio emission of \iras. 
The colours are the same as in Figure~\ref{fig:MSX}, with the red and green levels adjusted to bring out specific features observed within the bubble. 
Specifically the 24~\mum levels are now from 6 to 237.5~MJy/sr, whilst the 8~\mum emission is from 28 to 154~MJy/sr.
The radio contours are the same 1844~MHz ATCA contours as in Figure~\ref{fig:MSX}.
}
\label{fig:MSX_lobes}
\end{figure}

Figure~\ref{fig:MSX_lobes} shows a closeup of the radio bubble-structures of \iras.
The levels in this figure have been adjusted to bring out the detail in the emission in the bubble, which was saturated in Figure~\ref{fig:MSX}, so as to illustrate the low-surface-brightness and filamentary nature of the PAH emission surrounding the object.
Figure~\ref{fig:MSX_lobes} shows that the southeast bubble is dominated by dim emission from hot dust, which is surrounded by PAH emission, particularly towards the centre of the source, where the filled bubbles are observed to pinch together in the radio continuum.
This behaviour is also seen in the western bubble, although it is smaller and the dust emission is brighter and dominated by a strong point source at 24~\mum.
The central regions of \iras again show dust/PAH dominated emission, which is aligned in a direction perpendicular to the extension of the diffuse radio bubble. 

\begin{figure}
\centering
\includegraphics[width=0.55\textwidth]{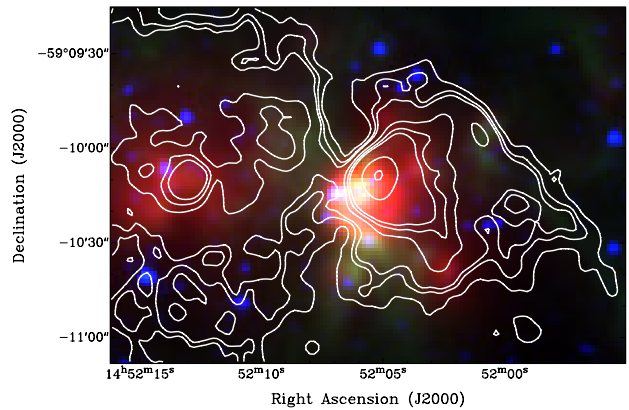}
\caption{Closeup of the central region of \iras. 
The colours are the same as in Figure~\ref{fig:MSX}, with the levels adjusted to bring out specific features observed within central parts of the source. 
Specifically the 24~\mum levels are now from 30 to 540~MJy/sr, whilst the 8 and 3.6~\mum emission runs from 25 to 550~MJy/sr and 20 to 40~MJy/sr, respectively.
The (white) radio contours are the same 1844~MHz ATCA contours as in Figure~\ref{fig:MSX}.
}
\label{fig:MSX_jet}
\end{figure}

Figure~\ref{fig:MSX_jet} shows a closeup of the central regions of \iras.
It reveals that at the point where the radio bubbles pinch and meet is a blue, compact source dominated by 3.6~\mum emission, suggesting that this is a young stellar object (YSO).
This object appears to be surrounded by a cocoon of hot dust emission with at least one linear feature, reminiscent of an outflow that is yellow in colour, suggesting that it results primarily from PAH and hot dust emission.
It is interesting, and maybe significant, that this outflow-like structure is pointing towards the strongest peak of radio emission in the western bubble.
A commensurate feature is not observed, however, towards the eastern bubble, which exhibits very low-surface-brightness PAH emission, again in the direction perpendicular to the bubble-structures observed in our radio continuum data.

Figure~\ref{fig:GLIMPSEoutflow} shows the region at 4.5~\mum, thought to be due to shocked molecular hydrogen and CO band heads, which possibly indicates the presence of an outflow \citep{Hindson2012}.
It is instructive, then, that the 4.5 ~\mum emission from this region is bright, and concentrated near the central compact object (seen as the bright blue emission in Figure~\ref{fig:MSX_jet}) and the brightest of the radio continuum emission regions.
This perhaps demonstrates that this object possesses an outflow \citep{Cyganowski2008b}.

Table~\ref{table:FIR} summarises the flux densities in the FIR waveband derived by the integration within the lowest contour of the radio emission at 1844~MHz.
We find flux densities of 15, 10, 27 and 43~Jy at the A, C, E and D bands of MSX.

\begin{deluxetable}{crr}
\tablewidth{0pt}
\tablecaption{Observed parameters of the sources within \iras in the FIR.\label{table:FIR}}
\tablehead{
\colhead{Frequency}           & \colhead{Average Flux Density}      & \colhead{Integrated Flux Density} \\
\colhead{(\mum)}		  & \colhead{(Jy/sr)}				   & \colhead{(Jy)} \\
\colhead{(1)}			  & \colhead{(2)}				   & \colhead{(3)}
} 
\startdata
3.6$^a$ & 6.3 & 38  \\
4.5$^a$ & 4.9 & 29 \\
5.8$^a$ & 26.4 & 158 \\
8.0$^a$ & 73.7 & 443 \\
8.28$^b$ & 14.8 & 15 \\
12.0$^c$  & 30.7 & 31 \\
12.13$^b$ & 9.9 & 10 \\
14.65$^b$ & 26.7 & 27 \\
21.34$^b$ & 42.7 & 43 \\
25$^c$  & 53.7 & 54 \\
60$^c$  & 192 & 470 \\
100$^c$  & 1302 & 1315 \\
\enddata
\tablenotetext{a}{IRAC/GLIMPSE data}
\tablenotetext{b}{MSX}
\tablenotetext{c}{IRAS}
\end{deluxetable}

\section{Discussion}
\subsection{Spectral Energy Distribution}
Figure~\ref{fig:SED} shows the spectral energy distribution of \iras from 843~MHz up to 3.6~\mum.
The SED shows that the emission at FIR frequencies long-ward of $\sim8$~\mum is due to thermal dust emission, whilst the radio continuum emission is well fitted with an optically thin thermal bremsstrahlung model.
The FIR and dust emission were fitted with a two-component ``greybody'' model, which is a modified blackbody curve, where the opacity is assumed to vary with frequency as $\tau_\nu=(\nu/\nu_0)^\beta$, where $\nu_0$ is the frequency at which the optical depth of that component is unity.
The flux density is then modelled according to $S_\nu=B_\nu(T_d)[1-\exp(-\tau_\nu)]\Omega_S$, where $B_\nu(T_d)$ is the Planck function at the dust temperature, $T_d$, and $\Omega_S$ is the solid angle subtended by the dust emitting part of the source (i.e., the region interior to the PAH emission in Figure~\ref{fig:MSX} and commensurate with the lowest radio continuum contour).
In the fitting, we have used an opacity index, $\beta$ of 2.0, following \citet{Garay2002} for sources with limited spectral points and consistent for sources showing signs of high-mass star formation \citep{Ossenkopf1994}.
We find a cold dust component at 19~K for an angular size of $3'\times4'$, assuming a Gaussian flux distribution within the source, and an optical depth of unity at 27~\mum.
The hot dust component is well-fitted with a temperature of 110~K.

The radio continuum emission below $\sim100$~GHz is well fit with an optically thin free-free emission, and to fit the emission, we used a technique of ``bootstrapping'' the thermal flux density  (\citet{Crocker2010}, Supplementary information) to that given by the 4500~MHz \pmn datum, F(4500~MHz), and assuming that it is completely thermal, so that the optical depth at 4500~MHz, $\tau(\textrm{4500~MHz})$ is given by:
\begin{equation}
	\tau(\textrm{4500~MHz})=\frac{F(\textrm{4500~MHz)}}{\Omega_SB(\textrm{4500~MHz},T)},
\end{equation}
where $\Omega_S$ is the solid angle of the source, and $B(\textrm{4500~MHz},T)$ is the Planck function at temperature, $T$.
We then obtain the optical depth at frequency $\nu$ from:
\begin{equation}
	\tau_\nu = \tau(\textrm{4500~MHz})\left(\frac{\nu}{10\textrm{~GHz}}\right)^{-2.1}.
\end{equation}
Hence the flux density, $S_\nu$ at any frequency is:
\begin{equation}
	S_\nu = \Omega_SB(\textrm{4500~MHz},T)[1-e^{-\tau_\nu}].
\end{equation}
Using a nominal electron temperature of 10,000~K, we fit the thermal emission.
The total luminosity radiated by from the source as a whole, found from integrating the emission shown in Figure~\ref{fig:SED} from 10~MHz up to 1~THz is $\sim3\times10^5$~L$_\odot$.

\subsection{Derived Parameters}
\subsubsection{Mass Estimate from Dust Emission}
Following \citet{Garay2002}, the fact that the thermal dust emission at 150~GHz is optically thin implies that the mass of the cloud, $M_g$ can be estimated:
\begin{equation}
	M_g=\frac{S_\nu D^2}{R_{dg}\kappa_\nu B_\nu(T_d)},
\end{equation}
where $\kappa_{\nu}$ is the mass absorption coefficient of the dust, $R_{dg}$ is the dust-to-gas ratio (assuming 10\% Helium), and $B_\nu(T_d)$ is the Planck function at the dust temperature, $T_d$.
Using the same values for the above equation as found in \cite{Garay2002}, we arrive at a mass estimate for \iras, parameterised as a function of the assumed distance of 12.7~kpc (as suggested by the \CO analysis in Section~\ref{sec:CO}) of:
\begin{equation}
	M_g\sim1.3\times10^4\left(\frac{D}{12.7\textrm{~kpc}}\right)^2\textrm{~M}_\odot.
\end{equation}
It must be explicitly noted, however, that even assuming a distance of 12.7~kpc, the above determination is dependant on $R_{dg}\kappa_\nu$, which is highly uncertain \citep{Garay2002}.
Hence we have also obtained mass estimates for \sourceIRAS from the \CO data. 

\subsubsection{ \CO and \isoCO Optical Depth Analysis and Mass Estimates}
From the fitting of the \CO and \isoCO line presented in Section~\ref{sec:CO}, we obtain an estimate of the optical depth of the \isoCO line from:
\begin{equation}
\tau({\rm \isoCO})=\frac{T_A({\rm \isoCO})}{T_A({\rm \CO})}\sim0.3,
\end{equation}
where $T_A({\rm \isoCO})$ are the fitted brightness-temperatures obtained from fitting the line profiles presented in Figure~\ref{fig:CO_line}, and assumed to be representative of the cloud as a whole.
Assuming a ratio of Carbon 12 to its isotope of $X_{12/13}=54$ \citep{Burton2013}, we then obtain an optical depth for the main line of $\tau({\rm \CO}) = \tau({\rm \isoCO})X_{12/13}\sim14$.
This shows that the \CO line is optically thick, but that the \isoCO is optically thin for this cloud.
Using the empirical relation describing the CO-to-H$_2$ conversion factor known as the X-factor from \citet{Bolatto2013} of $X_{\CO}=2\times10^{20}$ \cmsqr~(K km/s)$^{-1}$, we find a mass for the cloud of $\sim2\times10^4$~\msun, similar to that derived from the dust flux densities in the previous section.
We note that a more detailed analysis, such as that laid out in \cite{Burton2013} (and references therein), or virial mass estimates using the line parameters described in Section~\ref{sec:CO}, also arrive at a similar mass.
This mass also suggests the reason that the HOP survey did not observe \nhthree emission towards this source.

\begin{figure}
\centering
\includegraphics[width=0.5\textwidth]{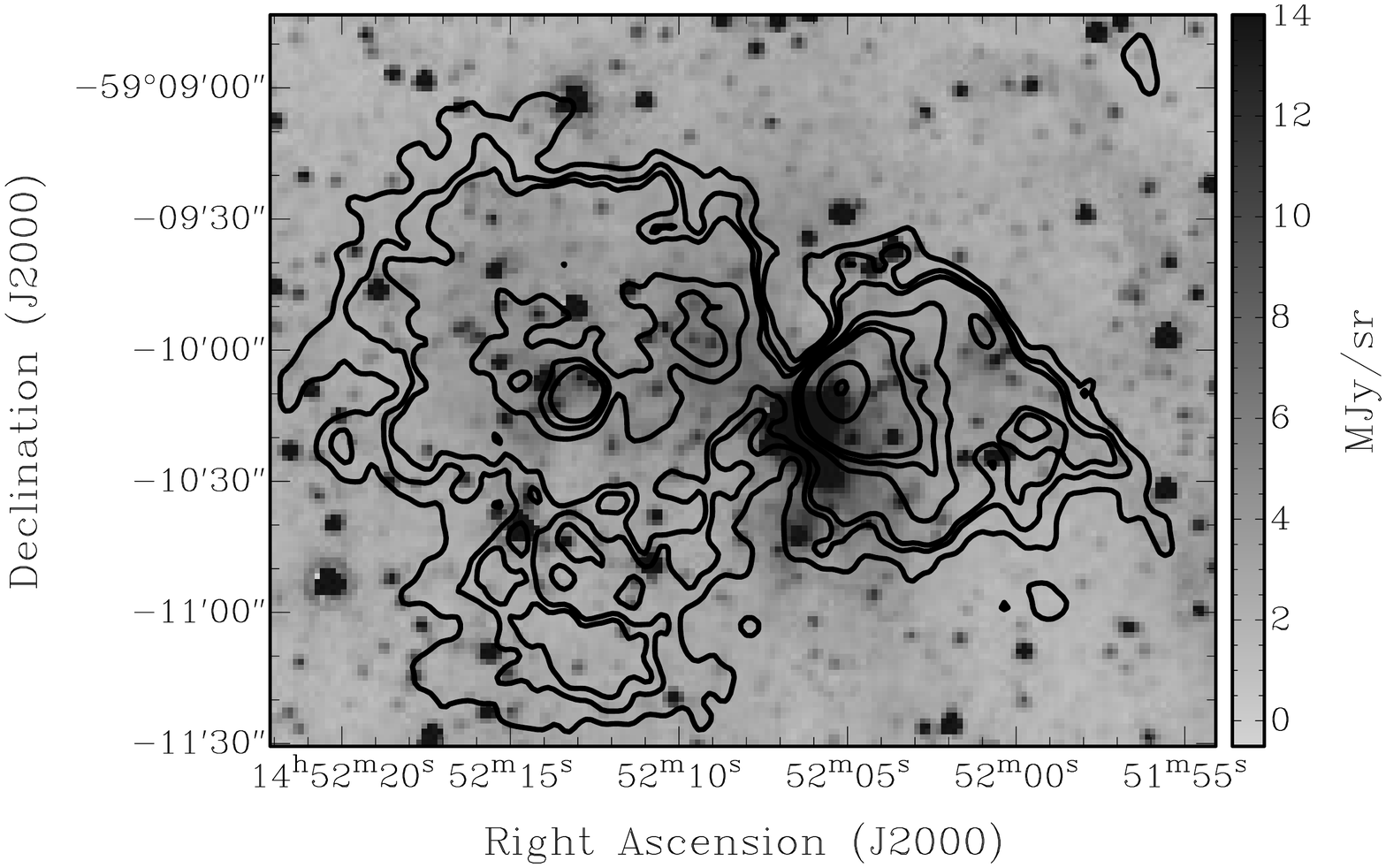}
\caption{A GLIMPSE 4.5~\mum image of IRAS~14482-5857, overlaid with same (white) ATCA contours as in Figure~\ref{fig:MSX}.
}
\label{fig:GLIMPSEoutflow}
\end{figure}

\subsection{Radio and FIR-derived parameters}
Sections \ref{sec:CO} and \ref{sec:water} strongly suggest that the emission from this region comes from a distance of $\sim12.7$~kpc.
This implies spatial dimensions for \iras $r\sim11\times14.7$~pc for an angular source extent of $3'\times4'$.
Figures~\ref{fig:ATCA} and \ref{fig:ATCA_1332} and Table~\ref{table:PSs} also show that there are three bright peaks of radio emission embedded within \iras.
Here we derive simple parameters for these regions assuming (as the spectral index maps in Figure~\ref{fig:SPIN} and integrated flux densities suggest) that these regions are H{\sc ii} regions.
We also derive the same parameters for the region as a whole.

From the source spectrum derived in the previous section and shown in Figure~\ref{fig:SED}, we find that \iras has an emission measure of $\sim1.4\times10^5$~pc~cm$^{-6}$, an electron number density of $\sim96$~\cmcube, and an ionising photon production rate of $\sim7\times10^{47}$~s$^{-1}$.
Assuming that the region as a whole is excited by a single zero-age main sequence star, it would be of spectral type B0.5.
Table~\ref{table:UCHIIregions} shows the parameters derived for each of the compact sources within \iras.
This shows that \iras consists of three UCH{\sc ii} regions (the bright peak sources A, B, and C) embedded within an H{\sc ii} region.
It also shows that, if the components that can be classed as UCH{\sc ii} regions are excited by individual zero-age main sequence stars, that the rates of UV photons required to ionise them imply spectral types of said stars of at least B0/1, except for source B, which is strong and compact, and at this assumed distance implies a star of spectral type O8.

\begin{deluxetable}{llllll}
\tablewidth{0pt}
\tablecaption{Derived parameters of the UCH{\sc ii} regions.\label{table:UCHIIregions}}
\tablehead{
\colhead{Region} & \colhead{Size} & \colhead{EM} & \colhead{$n_e$} & \colhead{$N_{Ly}$} & \colhead{Spectral } \\
\colhead{--} & \colhead{(pc)} & \colhead{ (pc~cm$^{-6}$)} & \colhead{(\cmcube)} & \colhead{(s$^{-1}$)} & \colhead{Type} \\
\colhead{(1)} & \colhead{(2)} & \colhead{(3)} & \colhead{(4)} & \colhead{(5)} & \colhead{(6)}  
}
\startdata
All & $14.5\times11$ & $1.4\times10^5$ & $96$ & $7\times10^{47}$ & B0.5 \\
A & $0.95\times0.80$ & $5\times10^6$ & $2.3\times10^3$ & $3\times10^{46}$ & B0 \\
B & $0.48\times0.43$ & $3.5\times10^7$ & $8.8\times10^3$ & $5\times10^{48}$ & O8 \\
C & $1.2\times0.61$ & $4\times10^5$ & $7\times10^2$ & $2\times10^{47}$ & B0 \\
\enddata
\end{deluxetable}

\subsection{The Nature of \iras}
The nature of the \iras, based on the evidence elucidated above, are strongly suggestive of massive star formation.
\citet{Walsh1997} suggest that this source is an UCH{\sc ii} region based on the IRAS colours and the Wood-Churchwell relation.
However, several studies of 6.7~GHz methanol maser emission towards this source failed to find any emission down to a limiting flux of 300~mJy/beam (c.f. Section~\ref{sec:otherMasers}).
We suggest that this resulted in this source being overlooked as a site of star formation.
Evidence of \iras being a site of {\it massive} star formation is indicated by the following: 
\begin{enumerate}
\item \CO and \isoCO observations showing strong emission towards the peaks of the radio continuum, which are dominated by emission at a velocity centroid of $V_{LSR}=-1.1$~km/s, with the \isoCO line not being able to be fitted for the velocity centroids at $V_{LSR}=-41.6$ and $-50.2$~km/s at all (though we note that this is probably due to the noise in the \isoCO cubes.
The derived (hydrogen) column density and mass (viz. $N_{H_2}\sim1.5\times10^{21}$~\cmsqr $\sim2\times10^4$~\msun) are commensurate with massive star forming regions.

\begin{figure}
\centering
\includegraphics[width=0.5\textwidth]{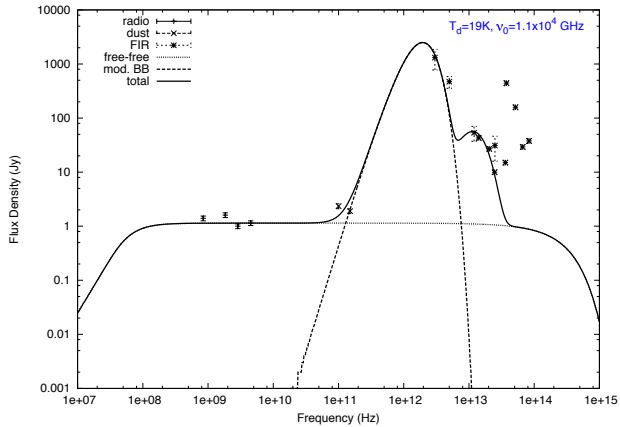}
\caption{Total spectral energy distribution (SED; solid black line) of \iras from radio continuum through to the IR.
Plus signs mark the radio (843 to 4500~MHz), crosses the dust emission at 100 and 150~GHz from \citet{QUaDsurveyB} and stars mark the IRAS, MSX and GLIMPSE fluxes as described in Table~\ref{table:FIR}.
The dotted curve is optically thin thermal bremsstrahlung emission modelled by ``bootstrapping'' the flux to the 4500~MHz PMN datum (see text for more details).
The dashed curve is a fit to the spectrum using two modified blackbody functions of the form $B_\nu(T_d){1-\exp[-(\nu/\nu_0)^\beta]}$, with different temperatures.
The fit parameters for the colder component are shown in the top right-hand corner.
}
\label{fig:SED}
\end{figure}

\item The presence of \water maser emission towards this source (c.f. Section~\ref{sec:water}).
The presence of the strong \CO emission at a velocity centroid of $V_{LSR}=-1.1$~km/s suggests that the velocity of the maser is not significantly displaced from velocity of the star-forming region itself.

\item A dust-derived cloud mass and temperature of $\sim10^5$~M$_\odot$ and 110~K, respectively.

\item The implied size, emission measure, electron number density and ionising photon production rate derived from the strong, compact radio continuum sources being representative of UCH{\sc ii} regions ionised by O/B stars.

\item The association of the brightest UCH{\sc ii} region with a bright 4.5~\mum source, which may be producing outflow-like structures observed at 8 and 24~\mum.
\end{enumerate}

We note that the image in Figure~\ref{fig:GLIMPSEoutflow} is subject to projection effects that may mimic such an outflow-like structure, so that this feature should be investigated further, but is nevertheless interesting.
The UCH{\sc ii} regions are surrounded by diffuse radio emission that itself can be considered an H{\sc ii} region.
This suggests, along with the bi-lobal radio structure that is seen perpendicular to the strong 8 and 24~\mum emission, that this source is a site of massive star formation at an advanced stage of star formation.
One possible objection to this hypothesis is that there are no class~{\sc i} methanol masers, which are known to be exclusively associated with high mass star formation \citep{Breen2013}.
However, the well known star forming region, Orion Source~I, which is forming massive stars, does not contain methanol maser emission \citep{Goddi2009,Matthews2010}

\section{Conclusions}
\iras has been the subject of few searches for southern sites of star formation even though it possessed IRAS colours suggestive of star formation, as the lack of 6.7~GHz methanol maser emission (down to a limiting flux density of 300~mJy/beam) meant that this source was not investigated further.
We have presented here new evidence that indeed \iras is a site of massive star formation.
Our results can be summarised as follows:
\begin{enumerate}
\item We have presented new, serendipitous ATCA observations of \iras at 1--3~GHz that are an order of magnitude higher in resolution ($\sim10''$) and sensitivity ($\sim100$~$\mu$Jy/beam) than available archival data.
These observations reveal that this source possesses a complex, bi-lobal morphology that is, spectrally, dominated by optically thin free-free emission in the broad, low-surface-brightness bubbles, strongly suggestive of an extended, classical H{\sc ii} region.
Embedded within this extended envelope are three compact sources that are suggestive of UCH{\sc ii} regions.

\item Spurred by the complex radio morphology, we have performed a detailed literature search and assembled a data-set that includes infrared data from 3.5 to 100~\mum, radio continuum to millimetre radio data from 843~MHz to 150~GHz, as well as maser and molecular line data from the HOPS and Mopra Southern Galactic Plane CO Surveys

\item The molecular line data, principally the Mopra Southern Galactic Plane CO Survey data, reveals a complex morphological structure.
Towards the brightest radio peak, the spectrum of the \CO emission exhibits three peaks, two near $V_{LSR}\sim-42$~km/s, and one near $V_{LSR}\sim0$~km/s.
The -- albeit noisier -- \isoCO spectrum, however, exhibits only a single peak structure centred near the $V_{LSR}\sim0$~km/s component of the \CO spectrum.

\item Fitting the \CO and \isoCO spectra, we find that the (dimmer) peak at $\sim-42$~km/s line emission can be fit by a two-component Gaussian, with velocity centroids at $V_{LSR}=-41.6\pm0.2$~km/s and $V_{LSR}=-50.2\pm0.2$~km/s. 
These peaks possess a brightness temperature of $T_{mb}=4.0\pm0.1$~K and a full-width at half-maximum (FWHM) of $\sigma_{FWHM}=1.3\pm0.1$~km/s for the former, and $T_{mb}=3.1\pm0.2$~K and $\sigma_{FWHM}=1.5\pm0.1$~km/s for the latter peak.
We find a bright peak of \CO data located at $V_{LSR}=-1.1\pm0.1$~km/s that is well fit using a brightness temperature of $T_{mb}=5.2\pm0.2$~K, and a FWHM width of $\sigma_{FWHM}=1.5\pm0.1$~km/s.
The \CO isotopologue \isoCO, also reveals line emission that is well-fit to $T_{mb}\sim1.4\pm0.1$~K at a velocity centroid of $V_{LSR}=-0.9\pm0.1$~km/s and a FWHM width of $\sigma_{FWHM}=1.0\pm0.1$~km/s, suggesting that the bulk of the mass is indeed at the farther distance.
We then used a water maser catalogued with the HOP survey that is positionally coincident with \iras, with a peak flux density of 5.1~Jy at --3.3~km/s with a line-width of 1.0~km/s.
We thus take the cloud to be at a velocity of $V_{LSR}=-1.1$~km/s, which places it at a distance of $\sim12.7$~kpc.

\item Using the \CO and \isoCO data, we have used two methods for estimating the hydrogen column density and mass towards \sourceIRAS, which we find give average values of $N_{H_2}\sim1\times10^{21}$~\cmsqr and $\sim2\times10^4$~\msun.

\item Using the distance estimate obtained from the \CO data, we thus fit sizes for the strong radio continuum sources of $\lesssim1$~pc, strongly suggestive of UCH{\sc ii} regions.
We then derive parameters for these sources: emission measures of $\sim10^{6}$~pc~cm$^{-6}$, electron number densities ($\sim10^3$~cm$^{-3}$), and rates of ionising photon production of $\sim10^{46-48}$~s$^{-1}$, consistent with OB stars.

\item 3-colour images of the region at 4.5, 8 and 24~\mum reveal a complex far-infrared (FIR) spectro-morphology.
The 4.5~\mum emission reveals a strong, compact source located at a minimum of the radio contours, indicative of a young stellar object.
Emanating from the point-source is what appears to be a outflow-like structure, which seems to align with the brightest of the three radio continuum peaks found in the object.
The point source and outflow-like structure appears to be surrounded by a region of hot dust, as indicated by the abundance of 24~\mum emission found there.
The emission within the diffuse 1--3~GHz radio contours is dominated by 24~\mum emission, suggesting that the radio bubbles are ``filled'' with hot dust.
Intersecting perpendicularly the lobular regions of hot dust is a region dominated by PAH emission, which also envelopes the diffuse radio emission with many wispy segments.

\item Spectrally, the region as a whole is well fit by an optically thin thermal bremsstrahlung and two-component modified blackbody spectrum.
The thermal bremsstrahlung component implies an emission measure of $\sim1.4\times10^{5}$~pc~cm$^{-6}$, an electron density of $\sim96$~cm$^{-3}$, a dust mass of $\sim6.701\times10^5$~M$_\odot$, and a production rate of ionising photons of $N_{Ly}$ of $\sim7\times10^{47}$~s$^{-1}$ for this source over its entire $3'\times4'$ extent.
The dust/FIR part of the spectrum is well fit by a two-component greybody with temperatures of $T_d=19$ and 110~K, respectively, strongly suggestive of massive star formation.

\end{enumerate}
In totality, the evidence -- distance estimates, strong radio sources that are consistent with UCH{\sc ii} regions, the presence of a water maser, a high mass derived from the \CO observations, and a high dust temperature -- strongly indicate that \sourceIRAS is a site of massive star formation.

\section*{Acknowledgments} We would like to thank the anonymous referee whose comments made the paper immeasurably better.
DIJ would like to thank Petter Hofverberg for assistance with the observations and Ciriaco Goddi and J\"{u}rgen Ott for enlightening discussions and a critical reading of the manuscript.
This research has made use of the SIMBAD database, operated at CDS, Strasbourg, France.
The Australia Telescope is funded by the Commonwealth of Australia for operation as a National Facility managed by CSIRO.

\bibliographystyle{apj} 
\bibliography{bibliographyPMN1452}

\end{document}